# "Model Cards for Model Reporting" in 2024: Reclassifying Category of Ethical Considerations in Terms of Trustworthiness and Risk Management

By DeBrae Kennedy-Mayo & Jake Gord[1]

**ABSTRACT**

In 2019, the paper entitled "Model Cards for Model Reporting" introduced a new tool for documenting model performance and encouraged the practice of transparent reporting for a defined list of categories [1]. One of the categories detailed in the 2019 paper is ethical considerations, which includes the subcategories of data, human life, mitigations, risks and harms, and use cases. We propose to reclassify this category in the original model card due to the recent maturing of the field known as trustworthy AI, a term which analyzes whether the algorithmic properties of the model indicate that the AI system is deserving of trust from its stakeholders. In our examination of trustworthy AI, we highlight three respected organizations - the European Commission's High-Level Expert Group on AI, the OECD, and the U.S.-based NIST - that have written guidelines on various aspects of trustworthy AI. These recent publications converge on numerous characteristics of the term, including accountability, explainability, fairness, privacy, reliability, robustness, safety, security, and transparency, while recognizing that the implementation of trustworthy AI varies by context. Our reclassification of the original model-card category known as ethical considerations involves a two-step process. First, we propose to add a new category known as trustworthiness, where the subcategories will be derived from the recent writings on trustworthy AI that we discuss in our paper. Second, instead of eliminating the subcategories of ethical considerations from the model card, we propose to maintain these subcategories under a renamed category known as risk environment and risk management, a title which we believe better captures today's understanding of the essence of these topics. We hope that this reclassification will further the goals of the original paper and continue to prompt those releasing trained models to accompany these models with documentation that will assist in the evaluation of the algorithmic properties of the models as the ethical, regulatory, and business demands concerning these models evolve.

## 1. INTRODUCTION

Today, countries around the world are grappling with how to appropriately regulate artificial intelligence (AI) [2, 3, 4].[2] In December 2023, the EU passed the first law to comprehensively regulate AI, the EU AI Act [5]. Although some worry that the recent urgency concerning AI may lead to potential regulatory solutions that are quickly crafted, it is important to realize that numerous organizations and experts have expended significant energies to help inform the discussion of responsible uses of AI in society. For example, the 2019 Ethics Guidelines for Trustworthy AI, crafted by the European Commission's High-Level Expert Group on Artificial Intelligence (EU HLEG), are viewed as the foundational underpinning of the EU's AI Act [6, 7]. In this paper, we examine three prominent guidelines – guidance from the EU HLEG, the OECD, and the U.S.-based NIST – which examine current concerns for AI under the term "trustworthy AI" [7, 8, 9]. Although these guidelines are not binding on companies, they can be viewed as soft law ("non-legislative policy instruments") that can assist organizations, in particular, in addressing complex issues related to AI that are not yet fully regulated [10, p. 389]. Noting the convergence among these guidelines in key areas, we revisit the model card, introduced in the 2019 paper entitled "Model Cards for Model Reporting," to update it with the lessons learned from trustworthy AI [1].

---

[1] DeBrae Kennedy-Mayo, Georgia Institute of Technology, debrae.kennedy-mayo@scheller.gatech.edu & Jake Gord, Georgia Institute of Technology, jgord3@gatech.edu

[2] In regulatory and governance efforts, the distinctions between AI, algorithms, models, and other related terms are often blurred. In this paper, we adopt the terms used by the relevant authors and organizations, instead of attempting to harmonize these terms with their technical meanings.

## 2. MODEL CARD FROM 2019

"Model Cards for Model Reporting" introduced a new tool for documentation of "trained machine learning (ML) and artificial intelligence (AI) models" and encouraged the practice of transparent reporting for a defined list of categories [1, p. 1]. The categories examined in the 2019 model card are "model details, intended use, factors, metrics, evaluation data, training data, quantitative analyses, ethical considerations, and caveats and recommendations" [1, p. 3]. Each of these categories is further divided into subcategories. The subcategories for the category of ethical considerations, for example, are: "data, human life, mitigations, risks and harms, and use cases" [1, p. 6].

The goal of the model card is to provide a framework to report "model provenance, usage, and ethics-informed evaluation" as well as to "give a detailed overview of a model's suggested uses and limitations that can benefit developers, regulators, and downstream users alike" [11]. While a model card provides insight concerning the assumptions underlying the model, the purpose of the model, the design process of the model, and the performance characteristics of the model, there is a concern that this type of documentation may be geared more towards practitioners in the field and regulators than to those affected by the model's outputs [12].

Although this paper will focus on the model card, it is worth pointing out that numerous alternative tools exist. Certification labels seek to distill characteristics concerning audited AI to a level that permits end users to make a determination about the trustworthiness of the AI, addressing a concern that end users lack the expertise to evaluate much of the material typically included in AI documentation [12, 13]. Ethics-based auditing can be used as a tool to examine how closely the practices of an organization comply with established ethics principles relevant to AI [6, 14]. Datasheets and data cards document information concerning the datasets that are used by the model, including topics such as what are the characteristics of the data set, how is the data collected, how is the processed, and what is the purpose for using the data [15, 16].

## 3. "ETHICAL CONSIDERATIONS" AND TRUSTWORTHINESS

In the five years since the publication of "Model Cards for Model Reporting," hundreds of academic articles have cited to the 2019 paper [1]. A subset of these academic articles has focused on ethics, trustworthiness, and the ethical considerations category from the model card, which is the category that is the focus of this paper [12, 14, 17, 18, 19, 20, 21, 22].

The description of the 2019 category of ethical considerations is as follows: "This section is intended to demonstrate the ethical consideration that went into model development, surfacing ethical challenges and solutions to stakeholders. Ethical analysis does not always lead to precise solutions but the process of ethical contemplation is worthwhile to inform on responsible practices and next steps in future work" [1, p. 6].

Although an in-depth study of AI ethics and trustworthiness is beyond the scope of this paper, we highlight similarities between these two topics. In our reclassification of the model card in Section 5, we propose intertwining "ethical considerations" and trustworthiness.

### 3.1 AI Ethics

AI ethics focuses on the development, deployment, and use of AI throughout its lifecycle [7]. As a field of study, AI ethics emerged in response to the challenges posed by AI such as "data bias, privacy, and fairness" [19, p. 2]. Commonly reported ethics failures related to AI are privacy intrusions (failure related to "consent to use data or consent to use data for intended purpose"), algorithmic bias[3] ("reaching a prediction that systematically disadvantages or even excludes one group based on personal identifiers"), and explainability (failure to "explain the decision that the AI algorithm has reached") [6, p. 56].

AI ethics can be divided into principles, which are often defined as "desirable properties of an AI system" [18, p. 700]. In 2018, Floridi and colleagues identified five ethical principles: "beneficence, non-maleficence, autonomy, justice, and explicability" [23, p. 696]. In 2019, Jobin and colleagues reported a convergence around the following principles: "transparency, justice and fairness, non-maleficence, responsibility, and privacy" [10, p. 3]. In 2023, Vainio-Pekka and colleagues noted seven principles associated with AI ethics: "transparency, responsibility, trust, privacy, sustainability, autonomy, and dignity" [19, p. 2]. Although these authors found convergence related to these ethical principles, divergence often existed in how the principles were interpreted and in which areas of concern manifested from the principles [10].

### 3.2 Trustworthy AI

Trustworthy AI is grounded in ethical principles [7, 18]. Trustworthy AI is a term that takes into account the stakeholders' expectations in an AI system and typically analyzes whether the algorithmic properties of the model indicate that the model is "developed, deployed, and used in a trustworthy manner," meaning that AI system is deserving of trust of these stakeholders [7, p. 10, 24].[4]

Trust is a complex concept which can be viewed through the lens of philosophy, psychology, sociology, and economics, with a core recurring theme of whether one entity placing confidence in another entity to act as expected is justified [25, 26]. From a practical perspective, trust can be described as "a confident relationship with the unknown," where the particular context ensures that stakeholders are able to bridge the gap between the known and the unknown [27].

In the context of AI, trust can then be viewed as "an attitude that an agent will help achieve an individual's goals in a situation characterized by uncertainty and vulnerability" [28, p.51]. In certain instances, the agent may act in a trustee role where "the trustee [is expected to] act in the trustor's best interests" [22, p. 625]. Both the developers of a model and the deployers of ta model can be envisioned as trustees [7].

---

[3] The "root cause" of algorithmic bias is most often attributable either to "programmers" or to "the veracity and relevance of the data" [6, p. 57].

[4] The term "stakeholders" is broadly defined to include both internal and external actors. Internal actors include C-suite, board members, scientists, engineers, and employees involved with risk and compliance. External actors. include "clients, vendors, customers, users, civil society, journalists, advocacy groups, community members, and others impacted by AI systems" [20, p. 10].

Generally speaking, the importance of trust is heightened in "high-stake AI scenarios" (which include "medical diagnosis, hiring procedure, and loan approval") as opposed to "low-stake AI scenarios" (which include "music preference, route planning, and price comparison") [29, p. 3, 12, p. 254].

**4. EU HLEG, OECD, AND NIST**

Three well-respected and influential organizations – the European Commission's High-Level Expert Group on AI (EU HLEG), the OECD, and the U.S.-based NIST – have written guidance on various aspects of trustworthy AI [7, 8, 9].[5] These recent publications converge on numerous characteristics of the term, including accountability, explainability, fairness, privacy, reliability, robustness, safety, security, and transparency.

Approaches to trustworthy AI are not a "one size fits all," but require detailed approaches that match with the types of risks presented, the types of data utilized by the model, and other factors. All of these guidelines note that the implementation of the characteristics of trustworthy AI vary by context [7, 8, 9].

The guidance from the three organizations also acknowledges that trustworthy AI is not accomplished by ensuring that all of the term's individual characteristics are incorporated, as not all characteristics apply in every setting. Instead, depending on the particular context, the importance of individual characteristics will vary. In addition, tradeoffs may be required among the characteristics [7, 8, 9, 25]. For example, requiring explainability in an AI system may negatively affect the accuracy of the system or may impact privacy or security [8]. Increasing the AI system's accuracy may occur at the cost of explainability [7]. As NIST states, "organizations can face difficult decisions in balancing these characteristics" [9, p. 12].

**4.1 Discussions of Trustworthiness**

The core characteristics for trustworthiness for the EU HLEG, the OECD, and the U.S.-based NIST are discussed below. A chart detailing the convergence of characteristics for these organizations is provided in Figure 1.

| AI CHARACTERISTICS | EU HLEG | OECD | US-based NIST |
|---|---|---|---|
| **Reliable & Robust** | Technical Robustness & Safety | Robustness, Security, & Safety | Valid & Reliable (includes Robust) |
| **Safety & Security** | Technical Robustness & Safety (includes Security) | Robustness, Security, & Safety | Safe<br>Secure & Resilient |
| **Privacy** | Privacy & Data Governance | Human-Centered Values (including Privacy) & Fairness | Privacy-Enhanced |
| **Transparent & Explainable** | Transparency (includes Explainability) | Transparency & Explainability | Accountable & Transparent<br>Explainable & Interpretable |

---

[5] In addition, numerous companies have published frameworks for trustworthy AI (or a comparable term), with similar characteristics identified. These companies include Google [30], IBM [31], Microsoft [32], Deloitte [33], and KPMG [34].

| **Fairness** | Diversity, Non-Discrimination, & Fairness | Human-Centered Values & Fairness | Fair & Bias Managed |
|---|---|---|---|
| **Accountable** | Accountability | Accountability | Accountable & Transparent |

**Figure 1: Convergence of Characteristics of Trustworthy AI from EU HLEG, OECD, and NIST**

**EU HLEG** [7]. In 2019, the EU High-Level Expert Group on Artificial Intelligence (EU HLEG) released its "Ethics Guidelines for Trustworthy AI" [6, 18, 35]. In the guidance, trustworthy AI has three elements: lawful (complying with existing laws and regulations), ethical (adhering to ethics-based values), and robust (performing as intended and addressing unintended harms). The overarching ethical principles for trustworthy AI are described as: "respect for human autonomy, prevention of harm, fairness, explicability" [7, p. 13]. From a practical perspective, these ethical principles can be implemented by enacting numerous characteristics of trustworthy AI: "human agency and oversight, technical robustness and safety, privacy and data governance, transparency, diversity, non-discrimination and fairness, societal and environmental wellbeing, and accountability" [7, p. 14-15]. For the EU AI Experts, security is an aspect of technical robustness and safety, while explainability is an aspect of transparency.

**OECD** [8, 36]. The Organisation for Economic Co-operation and Development (OECD) AI Principles "promote AI that is innovative and trustworthy and that respects human rights and democratic values" [37]. The 2019 OECD AI Principles detail the following characteristics: "inclusive growth, sustainable development, and well-being," "human-centered values and fairness," "transparency and explainability," "robustness, security, and safety," and "accountability" [8]. For the OECD, privacy is an aspect of human-centered values. The OECD does not explicitly mention reliability, but it can be inferred in the category of robustness, security, and safety.

**U.S. NIST** [9]. In 2023, the U.S. National Institute of Standards and Technology (NIST) published the "Artificial Intelligence Risk Management Framework" (AI RMF 1.0). This framework seeks both to minimize the negative risks of the AI system and to maximize positive impacts, noting that approaches which enhance trustworthy AI can assist in accomplishing these aims. AI RMF 1.0 identifies the following characteristics of trustworthy AI: "valid and reliable, safe, secure and resilient, accountable and transparent, explainable and interpretable, privacy-enhanced, and fair with harmful bias managed" [9, p. 12]. For NIST, robustness is an aspect of valid and reliable.

### 4.2 Convergence of Characteristics

The discussions of trustworthiness by the EU HLEG, the OECD, and the U.S.-based NIST converge on numerous characteristics of the term, including accountability, explainability, fairness, privacy, reliability, robustness, safety, security, and transparency, while recognizing that the implementation of trustworthy AI varies by context. Although these characteristics are interrelated and often discussed in tandem, each term is discussed individually in this section. As the section details the convergence in the publications examined, this is a non-exhaustive list of the characteristics of trustworthy AI.

*Accountability* - Accountability is anchored in an idea that an organization's actions will be linked to expectations, based on ethical, moral, or other guidance (such as an internal business practice or external code of conduct) [8]. Accountability can be envisioned as having numerous elements: "auditability" (assessment by internal and/or external auditors of data sets and processes that result in outcomes),

"minimization of negative impacts" (steps to ensure minimal undesirable outputs, such as risk management), "trade-offs" (addressing tensions that may arise among aspects of trustworthy AI by taking into account relevant values and interests), and "redress" (mechanisms that permit individuals to contest outputs of the model that affect them) [7, p. 19-20, 8, p. 15, 9, Principle 1.5]. Transparency is a pre-condition for accountability and is often necessary for appropriate redress [9].

*Explainability* – Explainability means enabling those directly (or indirectly) affected by a model output to understand the reasoning used by the model to generate a particular output. This reasoning can include the combination of the contributing input factors, the determinant factors, the data, and the logic [7, 8, 38]. In practice, providing this level of specificity concerning the model output in a manner that is understandable, by humans or machine-based analysis, may not be possible, meaning the decision cannot be adequately contested – a quandary which has been dubbed the "black box problem" [39, p. 117, 19, p. 6, 40]. In such instances, other measures may be needed to evaluate the model or system, such as "traceability" (documentation of data sets and processes utilized to yield outcomes) and "auditability" (assessment of data sets and processes that result in outcomes by internal and/or external auditors) [7, p. 13]. The importance of this concept is dependent on factors such as the context and the significance of the consequences if that output is deemed flawed [7, 8]. Explainability is related to interpretability (the meaning of the output in the context of the intended purposes) [9].

*Fairness* – Fairness addresses unfair bias and discrimination, with the implementation complicated because the views of these concepts vary across cultures [7, 9, 41, 42, 43, 44].[6] For the EU HLEG, the OECD, and the U.S.-based NIST, democratic values are at the core of this concept – such as freedom, equality, and social justice [7, 8, 9]. The concept of fairness can include both substantive and procedural aspects. The substantive aspect of fairness is linked to rights to non-discrimination, meaning that individuals and groups should be protected from "unfair bias, discrimination, and stigmatization" [7, p. 12]. The procedural aspect of fairness contemplates that individuals can contest decisions made by AI systems that affect them directly, and even indirectly.

*Privacy* – Privacy is a concept that focuses on "the norms and practices that help to safeguard human autonomy, identity, and dignity" [9, p. 17, 45]. Privacy protection requires appropriate data governance that covers the integrity and the quality of the data used, the model's access and processing of the data, and the data's relevance for the context in which the model is deployed [7, 46, 47]. Particular attention is warranted for data concerning individuals that is used to train the model as well as with data generated by the model that concerns specific individuals [7, 8, 9].

*Reliability* – A reliable model works as intended in a variety of situations when utilizing various types of inputs [7, 9].

*Robustness* – Robustness for the model, which can be termed technical robustness, means that the model performs as intended while minimizing unexpected harm if operating in unintended ways, particularly when misuse is foreseeable [7, 8, 9, 48]. Robustness can be viewed as developing the model to take a "preventative approach to risks" [7, p. 16].

---

[6] In this context, unfair bias can be difficult to define. Suresh and Guttag discuss numerous categories of bias that may have undesirable consequences: "historical bias," "representation bias", "measurement bias," "aggregation bias," "learning bias," "evaluation bias," and "deployment bias." [42, pp. 4-6].

*Safety* – The concept of safety means that the model is expected not to endanger humans when operating under expected conditions [8, 9]. Safety focuses on minimizing the occurrences of unintended consequences. Safety also includes designing processes to address these circumstances, such as requiring a review by a human. The appropriate level of safety measures is related to the risks posed by the model [7].

*Security* – The concept of security focuses on taking steps to prevent and mitigate abuses by malicious actors [7, 8]. Security is related to the concept of resiliency (the ability to have the AI model recover after an event by a malicious actor) [9].

*Transparency* – Transparency means providing individuals with information about the model (e.g. how the model is developed and trained) and how it arrives at its outputs, without the disclosure of proprietary code or datasets [8, 9]. Transparency can be conceived of having numerous elements: "traceability" (documenting the data sets and processes used to derive the model's outputs), "explainability" (ensuring that humans can understand the outputs of the model), and "communication" (informing humans that the model is being utilized in a particular situation) [7, p. 18]. Transparency can be conceived as a pre-condition to precisely determining whether the AI system is accurate, fair, secure, or enhances privacy [9].

**5. TWO-STEP RECLASSIFICATION**

After reviewing these guidelines on trustworthy AI in detail, we intend to incorporate key topics into the reclassification of the 2019 model-card category named "ethical considerations." This reclassification is a two-step process, with the intent to capture and further develop two differing aspects of the original category. In the place of the 2019 category of ethical considerations, the 2024 model card will have two categories: "trustworthiness" and "risk environment and risk management." Figure 2 provides a summary of the updates to the model card.

**2024 Updated Categories for Model Card**
- *Trustworthiness*
  – Reliable and Robust
  – Safety and Security
  – Privacy
  – Transparent and Explainable
  – Fair
  – Accountable
- *Risk Environment and Risk Management*
  – Data
  – Human Life
  – Risks and Harms
  – Mitigations
  – Use Cases

**Figure 2: Summary of Two-Step Reclassification of 2019 Category Known as Ethical Considerations**

### 5.1 Added category

The first step of the reclassification refocuses the 2019 category on the original description which states, in part, that the category is "intended to demonstrate the ethical considerations went into model development" and to note that "the process of ethical contemplation is worthwhile to inform on responsible choices" [1, p. 6]. Based on the recent developments in trustworthy AI, the 2024 model card adds a new category known as trustworthiness, with the new subcategories derived from the discussion above, to help to illuminate the ethical considerations undertaken concerning the model.

**Trustworthiness** – Trustworthy AI is a complex concept which will not manifest itself in the same way in every model. For each model, it is important to highlight which aspects have been important in the development of the model, and to address interactions between these aspects which lead to tradeoffs in the use of the model (e.g. the need for data privacy related to inputs may hamper the ability to explain outputs with the desired level of detail) [7, 9, 25].

**- Reliable and Robust** – Has the model been tested to ensure that it functions as expected? What steps have been taken to minimize harm if the model is misused in ways that are foreseeable?

**- Safety and Security** – What protections are in place to minimize unintended consequences of the model (e.g. human supervision)? What steps have been taken to prevent and mitigate abuses by malicious actors?

**- Privacy** – What steps have been taken to ensure appropriate practices related to the integrity and the quality of the data used, the model's access and processing of the data, and the data's relevance for the context in which the model is deployed? Are technical measures used to enhance privacy of data concerning individuals (e.g. anonymization, differential privacy, and encryption)?

**- Transparent and Explainable** – Can the model provide the reasons that it generated a particular output? If not, what other measures are in place (e.g. interpretability, auditability, or traceability) to evaluate the outputs of the model?

**- Fairness** – What strategies were used to evaluate and address unfair bias and discrimination resulting from the development and use of the model?

**- Accountable** – What steps have been taken to minimize negative impacts of the model? What mechanisms exist to permit an individual to contest a decision made by the model? Who is ultimately liable for any harm caused by leveraging a model?

Additional characteristics of trustworthy AI may be discussed, if appropriate.

### 5.2 Renamed category

In the second step of the reclassification, the subcategories of the 2019 category are maintained in the renamed category known as risk environment and risk management. Both of these topics focus on risk. Risk can be envisioned as the "effect of uncertainty" [24, p. 6]. The risk of a negative impact (for an individual, a group, or society) can be conceptualized as a combination of the likelihood of the outcome and the severity of the consequences of the outcome, if it were to occur [9]. Risk environment is intended to highlight aspects of the development or usage of the model which may indicate the need for higher levels of scrutiny for these aspects of the model. This topic can be associated with regulatory risk, reputational risk, and operational risk [6, 49]. Risk Management refers to an organization's efforts to coordinate activities to address risk, using approaches that can minimize potential negative impacts of the model (as well as to maximize positive impacts) [9, 50]. Another view of risk management is that it is designed to examine trustworthiness [24]. In higher risk environments, a full risk management assessment may be warranted. Although such an assessment is beyond the scope of the model card, the model card does provide a place to document that such an assessment exists.

**Risk Environment** examines two subcategories from the 2019 category of ethical considerations. The first of these is data. For data, a preliminary inquiry is whether the model uses data that is personal data (e.g. name, home address, email address) or that is linkable to a particular individual (e.g. postal code, approximate age, job title). If personal data is used, it is critical to know if the model makes use of sensitive data (or data related to protected classes). The second subcategory examines the types of outputs expected from the model. The inquiry is whether the model deals with decisions affecting human life, from regulating the proper functioning of a nuclear power plant to determining whether an individual is approved for a loan to purchase a house.

The term "risk environment" includes two subcategories from the original 2019 category of ethical considerations:

- "**Data** – Does the model use any sensitive data (e.g. protected classes)?" [1, p. 6].

- "**Human Life** – Is the model intended to inform decisions about matters central to human life or flourishing – e.g. health and safety? Or could it be used in such a way?" [1, p. 6].

**Risk Management** includes three subcategories from the 2019 category of ethical considerations. The first is risks and harms, which seeks to detail the risks and harms related to the model. In mitigations, the organization documents whether it has undertaken strategies to mitigate these risks. Use cases identifies any model uses that are likely to be problematic.

The term "risk management" includes three subcategories from the original 2019 category of ethical considerations:

- "**Risks and Harms** – What risks may be present in model usage? Try to identify the potential recipients, likelihood, and magnitude of harms. If these cannot be determined, note that they were considered but remained unknown" [1, p. 6].

- "**Mitigations** – What risk mitigation strategies were used during model development? [1, p. 6].

- "**Use Cases** – Are there any known model use cases which are particularly fraught? This may connect directly to the intended use section of the model card [1, p. 6].

## 6. DISCUSSION & FUTURE WORK

We have proposed an updated model card to capture both trustworthiness and risk management in the documentation provided by the card. We believe this emphasis is important for at least two reasons. First, trust is critical to the adoption of innovation as well as to the success of entrepreneurial ventures utilizing models [27]. Second, the updated model card recognizes the relationship of trust to risk. NIST explains the sentiment aptly by stating that "approaches which enhance AI trustworthiness can also contribute to a reduction of AI risks" [9, p. 12]. Risk based management can be leveraged by organizations to ensure the balance of trust-tradeoffs remains net positive throughout the model development lifecycle.

We believe that documenting an organization's efforts with respect to both trustworthiness and risk management are particularly important as the regulation of AI at the national and regional level is beginning to come online, such as with the passage of the EU AI Act in December 2023 [51]. The EU approach is risk based, meaning that the level of regulation varies depending on the risk posed. AI systems deemed to be high risk will be subject to significant regulation, while low risk AI systems will face minimum requirements [5, 52, 53]. The updated model card could prove beneficial in each of these scenarios. For low-risk AI systems, the model card can assist in providing stakeholders with detailed information about the model, even though such documentation is likely not required under the new law. Although the model card will not be sufficient to comply with the requirements of the new law in high-risk AI systems, the model card could be used to highlight aspects of trustworthiness to stakeholders. Given that compliance can also be driven at the commercial level, we believe that model cards can become a grassroots tool that allows organizations and their internal/external audit groups to police the sector well ahead of applicable legal requirements.

In future work, we hope to provide real-world examples of updated model cards driven by organizations wishing to communicate their AI trust posture to customers and end users. In addition, we plan to revisit the category of risk environment and risk management to explore the topic of societal externalities.

## 7. FUNDING

Partial funding for this paper was provided by ModelCard.ai.

**REFERENCES**

[1] Margaret Mitchell, Simone Wu, Andrew Zaldivar, Parker Barnes, Lucy Vasserman, Ben Hutchinson, Elena Spitzer, Inioluwa Deborah Raji, and Timnit Gebru. 2019. Model Cards for Model Reporting. *FAT* '19: Conference on Fairness, Accountability, and Transparency* (2019).
https://dl.acm.org/doi/10.1145/3287560.3287596.

[2] Blair Levin and Larry Downes. 2023. Who is Going to Regulate AI? *Harvard Business Review* (May 19, 2023). https://hbr.org/2023/05/who-is-going-to-regulate-ai.

[3] Dan Milmo and Kiran Stacey. 2023. Five Takeaways from UK's AI Safety Summit at Bletchley Park. *The Guardian* (November 2, 2023). https://www.theguardian.com/technology/2023/nov/02/five-takeaways-uk-ai-safety-summit-bletchley-park-rishi-sunak.

[4] Amanda Ruggeri. 2024. Davos 2024: Can and Should Leaders Aim to Regulate AI Directly? *BBC* (January 19, 2024). https://www.bbc.com/worklife/article/20240118-davos-2024-can-and-should-leaders-aim-to-regulate-ai-directly.

[5] Pamela Deese. 2023. EU Retains Role of Lead AI Regulator with Signing of EU AI Act. *The National Law Review* (December 18, 2023). https://www.natlawreview.com/article/eu-retains-role-lead-ai-regulator-signing-eu-ai-act.

[6] Luciano Floridi, Matthias Holweg, Mariarosaria Taddeo, Javier Amaya Silva, Jakob Mökander, and Yuni Wen. 2022. capAI - A Procedure for Conducting Conformity Assessment of AI Systems in Line with the EU Artificial Intelligence Act. *SSRN* (2022). https://dx.doi.org/10.2139/ssrn.4064091.

[7] Ethics Guidelines for Trustworthy AI. High-Level Expert Group on Artificial Intelligence. European Commission (2019). https://digital-strategy.ec.europa.eu/en/library/ethics-guidelines-trustworthy-ai.

[8] OECD Recommendation of the Council on Artificial Intelligence. OECD (2019). https://legalinstruments.oecd.org/en/instruments/OECD-LEGAL-0449.

[9] Artificial Intelligence Risk Management Framework (AI RMF 1.0). NIST (2023). https://nvlpubs.nist.gov/nistpubs/ai/NIST.AI.100-1.pdf.

[10] Anna Jobin, Marcello Ienca, and Effy Vayena. 2019. The Global Landscape of AI Ethics Guidelines. *Nature Machine Intelligence* (2019). https://www.nature.com/articles/s42256-019-0088-2.

[11] Huanming Fang and Hui Miao. 2020. Introducing the Model Card Toolkit for Easier Model Transparency Reporting. *Google Research Blog* (2020). https://blog.research.google/2020/07/introducing-model-card-toolkit-for.html.

[12] Nicolas Scharowski, Michaela Benk, Swen J. Kuhne, Leane Wettstein, and Florian Bruhlmann. 2023. Certification for Trustworthy AI: Insights from an Empirical Mixed-Method Study. *FAccT '23:, ACM Conference on Fairness, Accountability, and Transparency* (2023). https://arxiv.org/pdf/2305.18307.pdf.

[13] Kees Stuurman and Eric Lachaud. 2022. Regulating AI: A Label to Complete the Proposed Act on Artificial Intelligence. *Computer Law & Security Review* (2022). https://doi.org/10.1016/j.clsr.2022.105657.

[14] Jakob Mökander. 2023. Auditing of AI: Legal, Ethical, and Technical Approaches. *DISO* (2023). https://doi.org/10.1007/s44206-023-00074-y.

[15] Timnit Gebru, Jamie Morgenstern, Briana Vecchione, Jennifer Wortman Vaughan, Hanna Wallach, Hal Daumé III, and Kate Crawford. 2021. Datasheets for Datasets. *Commun. ACM* (2021). https://doi.org/10.1145/3458723.

[16] Mahima Pushkarna, Andrew Zaldivar, and Oddur Kjartansson. 2022. Data Cards: Purposeful and Transparent Dataset Documentation for Responsible AI. *FAccT '22: ACM conference on Fairness, Accountability, and Transparency* (2022). https://dl.acm.org/doi/10.1145/3531146.3533231.

[17] Michale Mylrea and Nikki Robinson. 2023. Artificial Intelligence (AI) Trust Framework and Maturity Model: Applying an Entropy Lens to Improve Security, Privacy, and Ethical AI. *Entropy*. (2023). https://www.mdpi.com/1099-4300/25/10/1429.

[18] Erich Prem. 2023. From Ethical AI Frameworks to Tools: A Review of Approaches. *AI Ethics* (2023). https://link.springer.com/article/10.1007/s43681-023-00258-9#citeas.

[19] Heidi Vainio-Pekka, Mamia Ori-Otse Agbese, Marianna Jantunen, Ville Vakkuri, Tommi Mikkonen, Rebekah Rousi, and Pekka Abrahamsson. 2023. The Role of Explainable AI in the Research Field of AI Ethics. *ACM Trans. Interact. Intell. Syst.* (2023). https://doi.org/10.1145/3599974.

[20] Richmond Wong, Michael Madaio, and Nick Merrill. 2023. Seeing Like a Toolkit: How Toolkits Envision the Work of AI Ethics. *ACM Hum.-Comp. Interact*. (2023). https://dl.acm.org/doi/pdf/10.1145/3579621.

[21] Jose Luiz Nunes, Gabriel D. J. Barbosa, Clarisse Sieckenius de Souza, Helio Lopes, and Simone D. J. Barbosa. 2022. Using Model Cards for Ethical Reflection: A Qualitative Exploration. *IHC '22: Brazilian Symposium on Human Factors in Computing Systems* (2022). https://dl.acm.org/doi/10.1145/3554364.3559117.

[22] Alon Jacovi, Ana Marasović, Tim Miller, and Yoav Goldberg. 2021. Formalizing Trust in Artificial Intelligence: Prerequisites, Causes and Goals of Human Trust in AI. *FAccT '21: ACM Conference on Fairness, Accountability, and Transparency* (2021). https://doi.org/10.1145/3442188.3445923.

[23] Luciano Floridi, Josh Cowls, Monica Beltrametti, Raja Chatila, Patrice Chazerand, Virginia Dignum, Christoph Luetge, Robert Madelin, Ugo Pagalio, Francesca Rossi, Burkhard Schafer, Peggy Valcke, and Effy Vayena. 2018. AI4People – An Ethical Framework for a Good AI Society: Opportunities, Risks, Principles, and Recommendations. *Minds and Machines* (2018). https://link.springer.com/article/10.1007/s11023-018-9482-5.

[24] Helen Smith, Arianna Manzini, Mari-Rose Kennedy, and Jonathan Ives. 2023. Ethics of Trust/worthiness in Autonomous Systems: A Scoping Review. *TAS '23: First International Symposium on Trustworthy Autonomous Systems* (2023). https://doi.org/10.1145/3597512.3600207.

[25] Bo Li, Peng Qi, Bo Liu, Shuai Di, Jingen Liu, Jiquan Pei, Jinfeng Yi, and Bowen Zhou. 2023. Trustworthy AI: From Principles to Practices. *ACM Computing Surveys* (2023). https://dl.acm.org/doi/full/10.1145/3555803.

[26] Davinder Kaur, Suleyman Uslu, Kaley Rittichier, and Arjan Durresi. 2022. Trustworthy Artificial Intelligence: A Review. *ACM Computing Surveys* (2022). https://dl.acm.org/doi/full/10.1145/3491209?casa_token=mLo7dXCJSdgAAAAA%3AxIkKctfRGiMlyxNXGRr30o3zIWYKP8cQICpyqmZ6dAA_yHogVQoNCLsIxaXVERa2M6ie0fvNb4N2#Bib0023.

[27] Rachel Botsman. 2017. Who Can You Trust? Public Affairs (2017).

[28] John D. Lee and Katrina A. See. Trust in Automation: Designing for Appropriate Reliance. *Human Factors* (2004). https://journals.sagepub.com/doi/abs/10.1518/hfes.46.1.50_30392.

[29] Shivani Kapania, Oliver Siy, Gabe Clapper, Azhagu Meena SP, and Nithya Sambasivan. 2022. Because AI is 100% Right and Safe: User Attitudes and Sources of AI Authority in India. *CHI '22: CHI Conference on Human Factors in Computing Systems* (2022). https://doi.org/10.1145/3491102.3517533.

[30] Responsibility: Our Principles. Google AI. https://ai.google/responsibility/principles/.

[31] Trustworthy AI. IBM. https://research.ibm.com/topics/trustworthy-ai.

[32] Responsible AI: Principles and Approach. Microsoft AI. https://www.microsoft.com/en-us/ai/principles-and-approach.

[33] Trustworthy AI: Bridging the Ethics Gap Surrounding AI. Deloitte. https://www2.deloitte.com/us/en/pages/deloitte-analytics/solutions/ethics-of-ai-framework.html.

[34] KPMG Trusted AI. KPMG. https://kpmg.com/us/en/careers-and-culture/trusted-ai.html?utm_source=bing&utm_medium=cpc&utm_campaign=7014W000001j9caQAA&cid=7014W000001j9caQA.

[35] Luciano Floridi. 2019. Establishing the Rules for Building Trustworthy AI. Nature Machine Intelligence (2019). https://www.nature.com/articles/s42256-019-0055-y.

[36] OECD AI Principles Overview. OECD (2019). https://oecd.ai/en/ai-principles.

[37] Artificial Intelligence: OECD Principles: How Governments and Other Actors Can Shape a Human-Centric Approach to Trustworthy AI. OECD (undated). https://www.oecd.org/digital/artificial-intelligence/#:~:text=The%20OECD%20Principles%20on%20Artificial%20Intelligence%20promote%20AI,approved%20the%20OECD%20Council%20Recommendation%20on%20Artificial%20Intelligence.

[38] Olivia Tracey and Robert Irish. 2023. Explainability for All: Care Ethics for Implementing Artificial Intelligence. *ITSAS: IEEE International Symposium on Technology and Society* (2023). https://ieeexplore.ieee.org/stamp/stamp.jsp?tp=&arnumber=10306199.

[39] Mark Coeckelbergh, AI ETHICS, MIT Press (2020).

[40] Deven Desai and Joshua Kroll. 2017. Trust but Verify: A Guide to Algorithms and the Law. *Harvard Journal of Law and Technology* (2017). https://papers.ssrn.com/sol3/papers.cfm?abstract_id=2959472.

[41] Agathe Balayn, Christoph Lofi, and Geert-Jan Houben. 2021. Managing Bias and Unfairness in Data for Decision Support: A Survey of Machine Learning and Data Engineering Approaches to Identify and Mitigate Bias and Unfairness within Data Management and Analytics Systems. *VLDB Journal* (2021). https://link.springer.com/article/10.1007/s00778-021-00671-8.

[42] Harini Suresh and John Guttag. 2021. A Framework for Understanding Unintended Consequences of Machine Learning. *EAAMO '21: Equality and Access in Algorithms, Mechanisms, and Optimization* (2021). https://arxiv.org/pdf/1901.10002.pdf.

[43] Michael Kerans and Aaron Roth. 2020. Chapter: Algorithmic Fairness - From Parity to Pareto. THE ETHICAL ALGORITHM: THE SCIENCE OF SOCIALLY AWARE ALGORITHM DESIGN. Oxford University Press (2020).

[44] Hannah Fry. 2018. Chapter: Justice, HELLO WORLD: BEING HUMAN IN THE AGE OF ALGORITHMS. Norton (2018).

[45] Peter Swire and DeBrae Kennedy-Mayo. 2024. Chapter 1: Introduction to Privacy. U.S. PRIVATE-SECTOR PRIVACY. International Association of Privacy Professionals (2024).

[46] Michael Kerans and Aaron Roth. 2020. Chapter: Algorithmic Privacy - From Anonymity to Noise. THE ETHICAL ALGORITHM: THE SCIENCE OF SOCIALLY AWARE ALGORITHM DESIGN. Oxford University Press (2020).

[47] Hao-Ping, Yu-Ju Yang, Thomas Serban Von Davier, Jodi Forlizzi, and Sauvik Das. 2023. Deep Fakes, Phrenology, Surveillance, and More! A Taxonomy of Privacy Risks. *Arvix* (2023). https://arxiv.org/pdf/2310.07879.pdf.

[48] Subhabrata Majumdar. 2023. Chapter 3: Fairness, Explainability, Privacy, and Robustness for Trustworthy Algorithmic Decision-Making. BIG DATA ANALYTICS IN CHEMOINFORMATICS AND BIOINFORMATICS.

Elsevier (2023). https://www.sciencedirect.com/science/article/pii/B9780323857130000177?via%3Dihub.

[49] Richard Wong, Andrew Chong, and R. Cooper Aspergren. 2023. Privacy Legislation as Business Risks: How GDPR and CCPA are Represented in Technology Companies' Investment Risk Disclosures. *Proc. ACM Hum.-Comp. Interact*. (2023). https://dl.acm.org/doi/pdf/10.1145/3579515.

[50] Andre Steimers and Moritz Schneider. 2022. Sources of Risks of AI Systems. *Int. J. Environ. Res. Public Health* (2022). https://www.ncbi.nlm.nih.gov/pmc/articles/PMC8951316/.

[51] Artificial Intelligence Act: Deal on Comprehensive Rules for Trustworthy AI. European Parliament (December 9, 2023). https://www.europarl.europa.eu/news/en/press-room/20231206IPR15699/artificial-intelligence-act-deal-on-comprehensive-rules-for-trustworthy-ai.

[52] Melissa Heikkila. 2023. Five Things You Need to Know about the EU's New AI Act. *MIT Technology Review* (December 11, 2023). https://www.technologyreview.com/2023/12/11/1084942/five-things-you-need-to-know-about-the-eus-new-ai-act/.

[53] Kees Stuurman and Eric Lachaud. 2022. Regulating AI: A Label to Complete the Proposed Act on Artificial Intelligence. *Computer Law & Security Review* (2022). https://doi.org/10.1016/j.clsr.2022.105657.